\documentstyle[12pt]{article}
\newcommand{\beq}{\begin{equation}}
\newcommand{\eeq}{\end{equation}}
\newcommand{\ba}{\begin{eqnarray}}
\newcommand{\ea}{\end{eqnarray}}

\newcommand{\dsl}
  {\kern.06em\hbox{\raise.15ex\hbox{$/$}\kern-.56em\hbox{$\partial$}}}

\newcommand{\eeqarr}{\end{eqnarray}}
\newcommand{\ZZ}{{\rm \kern 0.275em Z \kern -0.92em Z}\;}
\begin{document}
\begin{titlepage}
\begin{center}
{\Huge Bound States in the AdS/CFT}
\\
\vspace*{0.1cm}
{\Huge Correspondence}
\\
\vspace*{0.7cm}
{\large Pablo Minces\footnote{pablo@fma.if.usp.br}}
\\
\vspace*{0.3cm}
Centro Brasileiro de Pesquisas F\'{\i}sicas (CBPF),\\
Departamento
de Teoria de Campos e Part\'{\i}culas (DCP),\\ Rua Dr. Xavier Sigaud 150,
22290-180, Rio de Janeiro, RJ, Brazil.
\vspace*{0.7cm}

\end{center}
\begin{abstract}
We consider a massive scalar field theory in anti-de Sitter space, in 
both minimally and non-minimally coupled cases. We introduce 
a relevant double-trace perturbation at the boundary, by carefully 
identifying the correct source and generating functional for the 
corresponding conformal operator. We show 
that such a relevant double-trace perturbation introduces changes in the 
coefficients in the boundary terms of the action, 
which in turn govern the existence of a bound state in the bulk. For 
instance, 
we show that the usual action, containing no additional boundary terms, 
gives rise to a bound state, which can be avoided only through the 
addition of 
a proper boundary term. Another notorious example is that of a conformally 
coupled 
scalar field, supplemented by a Gibbons-Hawking term, for which there is 
no associated bound state. In 
general, in both minimally and non-minimally coupled cases, we explicitly 
compute the boundary terms which give rise to a bound 
state, and which ones do not. In the non-minimally coupled case, and 
when the action is supplemented by a Gibbons-Hawking term, this also 
fixes allowed values of the coupling coefficient to the 
metric. We interpret our results as the fact that the requirement to 
satisfy the
Breitenlohner-Freedman bound does not suffice to prevent tachyonic 
behavior 
from existing in the bulk, as it must 
be supplemented by additional conditions on the coefficients in 
the boundary terms of the action.
\end{abstract}

\vskip 0.7cm

\begin{flushleft}
\end{flushleft}
\end{titlepage}

\section{Introduction}

The AdS/CFT correspondence \cite{maldacena}\cite{witten}\cite{gubser} (see 
\cite{maldacena2}\cite{freedman4} for reviews) proposes the existence of a 
duality between a field theory on ($d+1$)-dimensional anti-de Sitter 
($AdS_{d+1}$) space, and a 
Conformal Field Theory (CFT) living at its boundary, and 
since its 
formulation, a large amount of work has been devoted to explore different 
aspects of this conjecture. A prescription for mapping one theory into the 
other was proposed in \cite{witten}\cite{gubser}, and it reads

\beq
\exp\left(-I_{AdS}[\phi_{0}]\right)\equiv 
\left<\exp\left(\int d^{d}x\; {\cal O}(\vec{x})\phi 
_{0}(\vec{x})\right)\right>\; ,
\label{9003}
\eeq
where $\phi_{o}$ is the
boundary value of the bulk field $\phi$, and it couples to the 
boundary CFT operator ${\cal O}$. Throughout this paper, we will be 
concerned with a scalar field theory in AdS. In the minimally coupled 
case, the action reads

\beq
I_{0}=-\frac{1}{2}\;\int d^{d+1} y \;\sqrt{g}\;
\left(
g^{\mu\nu}\partial_{\mu}\phi\;\partial_{\nu}\phi\; +
\;m^{2}\phi^{2}\right) \; ,
\label{301}
\eeq
where $m$ is the mass of the scalar field. The corresponding equation of 
motion is of the form $(\nabla^{2} - m^{2})\phi =0$.

One relevant aspect of the study of the AdS/CFT conjecture is the analysis 
of perturbations of the boundary CFT by double-trace operators. It was 
proposed in \cite{berkooz1}\cite{berkooz3} that they give rise to a new 
perturbation expansion for string theory, based on a non-local 
world-sheet. Later, it was suggested in \cite{witten7}\cite{berkooz2} that 
multi-trace interactions can be incorporated in the AdS/CFT correspondence 
by generalizing the boundary conditions which are considered in the usual 
single-trace case.

In any prescription describing such phenomenon, we should take into 
account 
the existence of 
two normalizable modes for the scalar field on AdS 
\cite{freedman}\cite{freedman8} (see also
\cite{mezincescu}), namely `regular' and 
`irregular' ones, which behave close to the border as 
$\phi_{R}\sim\epsilon^{\Delta_{+}}$, for regular modes, and 
$\phi_{I}\sim\epsilon^{\Delta_{-}}$ for irregular ones. Here $\epsilon$ 
is a measure of the distance to the
boundary, and

\beq
\Delta_{\pm}=\frac{d}{2}\;\pm\;\nu\; ,
\label{9006''}
\eeq
\beq
\nu=\sqrt{\frac{d^2}{4}\;+\;m^{2}}\; ,
\label{9006'''}
\eeq
where $m$ satisfies the Breitenlohner-Freedman bound 

\beq
m^{2}\geq -\frac{d^2}{4}\; .
\label{9034}
\eeq
The range $m^{2}<-\frac{d^2}{4}$ corresponds 
to tachyons in AdS \cite{freedman}\cite{freedman8}, and, in fact, if 
Eq.(\ref{9034}) is not satisfied, the energy is neither conserved nor 
positive definite \cite{mezincescu}. It was also shown in 
\cite{freedman}\cite{freedman8} that irregular modes are 
normalizable only for

\beq
\nu < 1\; .
\label{12000}
\eeq

In the AdS/CFT picture, the interpretation of the above results was 
considered in \cite{witten2} (see also \cite{kraus} for previous 
results), which points out that we should find two 
different CFT's at the boundary, corresponding to both possible 
quantizations in the bulk. 
However, the usual prescription Eq.(\ref{9003}) accounts for only the CFT 
with conformal dimension $\Delta_{+}$, corresponding to regular modes 
propagating in the bulk. In order to also reproduce the missing conformal 
dimension $\Delta_{-}$, corresponding to irregular modes in the bulk, the 
proposal in \cite{witten2} was that its generating functional could be 
found by performing a Legendre transformation to the original one 
in the theory with conformal dimension $\Delta_{+}$ (see also 
\cite{dobrev} for previous results involving group-theoretic analysis). 
Thus, starting from the generating functional in the theory with conformal 
dimension $\Delta_{+}$, as considered in 
\cite{witten}\cite{freedman3}\cite{viswa1}, it was explicitly shown in 
\cite{witten2} that its Legendre transform gives rise to the conformal 
dimension $\Delta_{-}$, as expected.

Note that, near the boundary, the scalar field behaves as (for $\nu<1$)

\beq
\phi(\epsilon,\vec{x})=\epsilon^{\Delta_{+}}\alpha(\vec{x})+
\epsilon^{\Delta_{-}}\beta(\vec{x})\; ,
\label{8000}
\eeq
where $\vec{x}$ are coordinates in the boundary. One possibility is to 
impose the boundary condition $\beta(\vec{x})=0$. In this circumstance,
$\alpha(\vec{x})$ is understood as the expectation value of a conformal 
operator ${\cal O}_{\beta}$ with dimension $\Delta_{+}$ 
\cite{witten2}\cite{kraus}. On the other hand, when we consider the 
boundary condition

\beq
\alpha(\vec{x})=0\; ,
\label{8006}
\eeq
then irregular modes, instead of regular ones, propagate in the bulk. Now 
the boundary theory has a conformal operator ${\cal O}_{\alpha}$ of 
dimension $\Delta_{-}$, whose expectation value is given by

\beq
\beta(\vec{x})\equiv\left<{\cal O}_{\alpha}(\vec{x})\right>\; .
\label{8007}
\eeq

In order to describe the way in which double-trace perturbations are
incorporated in the AdS/CFT conjecture, we first note that, since 
$2\Delta_{-}<d$, a relevant double-trace deformation should be of the form 
\cite{witten7}

\beq
W[{\cal O}_{\alpha}]=\frac{f}{2}\;{\cal O}_{\alpha}^{2}\; ,
\label{8001}
\eeq
where $f$ is a coupling constant, and, as pointed out before, ${\cal 
O}_{\alpha}$ has dimension $\Delta_{-}$. Then, the prescription in 
\cite{witten7} is that the above double-trace perturbation is 
described by 
the generalized boundary condition

\beq
\alpha=\;f\;\beta\; .
\label{8002}
\eeq
Note that, for $f=0$, the above boundary condition reduces to 
Eq.(\ref{8006}), as expected. The above equation describes a 
renormalization group (RG) flow \cite{witten7}, starting 
from 
the UV fixed point at $f=0$, 
and having an endpoint at an IR fixed point whose generating functional 
is related to the one of the $f=0$ case by a Legendre transformation, as 
explained above (see \cite{gubser3}\cite{gubser4}\cite{odintsov} for 
analyses on this 
subject). 
Additional references on the topic of double-trace interactions in 
the AdS/CFT correspondence are 
\cite{muck}\cite{our4}\cite{petkou}\cite{sever}\cite{barbon}\cite{berkooz8}\cite{witten5}\cite{troost}\cite{nolland}\cite{hoyos}. 

In particular, in this work we will be concerned with the recent results 
in 
\cite{troost}, regarding unstable double-trace perturbations. As pointed 
out in \cite{witten7}, stability of a double-trace deformation is 
related to the sign of its coefficient $f$ (see Eq.(\ref{8001})). 
Specifically, stable perturbations correspond to $f>0$, whereas 
unstable ones exist for $f<0$. The author of \cite{troost} 
asks how the bulk theory detects an unstable theory in the boundary, and, 
in order to answer such a question, considers a careful analysis of the 
solution to the radial wave-equation 
for a massive
scalar field in AdS. Calculations are performed in the 
representation of the $AdS_{d+1}$ space in Lorentzian Poincar\'e 
coordinates. The author of \cite{troost} points out the existence of a 
bound 
state 
having tachyonic behavior, in 
the spectrum of the radial equation, when $\alpha$ and $\beta$ 
in Eq.(\ref{8000}) satisfy

\beq
\frac{\alpha}{\beta}\;<0\; .
\label{8009}
\eeq
But, on the other hand, we note from Eq.(\ref{8002}) that the above 
condition is equivalent to set $f<0$, and this led the author of 
\cite{troost} to conclude that an unstable double-trace deformation in 
the boundary corresponds to the existence of a solution to the 
bulk wave-equation with tachyonic behavior in a Minkowski slice. Note 
that the tachyonic behavior appears even when the 
Breitenlohner-Freedman bound Eq.(\ref{9034}) is satisfied. This 
result adds a new entry to the AdS/CFT dictionary, and, as pointed out in 
\cite{troost}, could be relevant to the analysis of causality 
and Lorentzian 
aspects of the AdS/CFT correspondence.

One of the purposes of this paper is to propose a deeper 
insight into the above 
detailed results. In particular, we will be concerned with the role of the 
action in the phenomenon of the existence of bound states for the scalar 
field in AdS. Note that, as emphasized in \cite{troost}, a bulk theory is 
specified not only by the background geometry, but also by the boundary 
conditions which are imposed on the bulk field. But boundary conditions 
are governed by the action, and this suggests that a careful 
analysis, in this context, of the action 
of the bulk theory, could shed some new light into the phenomenon of the 
existence of bound states.

There have been previous situations where a careful study of the 
action, and not only of its corresponding equation of motion, has proven 
to be 
fruitful in the context of the AdS/CFT correspondence. An example of this 
is the case of the spinor field, whose action contains, at most, first 
order derivatives, and vanishes on-shell. Because of this, it was pointed 
out in \cite{sfetsos} that, in order to compute the generating functional 
for the 
corresponding dual CFT, a boundary term should be added to the bulk 
action. Such a boundary term was later computed using the Hamiltonian 
formalism \cite{frolov} and the Variational Principle \cite{henneaux}. 
Analogous situations were found in the cases of the antisymmetric tensor 
field \cite{frolov2} and the Self-Dual model \cite{our}.

In the case of the scalar field theory, it was shown in \cite{our2} that 
a boundary CFT with conformal dimension $\Delta_{-}$ could be generated by 
adding a proper boundary term to the usual action. Later, such result was 
combined with the Legendre transform prescription in \cite{witten2} in 
order to find a generalized AdS/CFT prescription which 
is able to map to the boundary all constraints arising from the 
quantization in the bulk. In order to illustrate this, we consider a 
non-minimally coupled scalar field, 
where the action Eq.(\ref{301}) should be replaced by

\beq
{\cal I}_{0}=-\frac{1}{2}\;\int d^{d+1} y \;\sqrt{g}\;   
\left[
g^{\mu\nu}\partial_{\mu}\phi\;\partial_{\nu}\phi\; +
\;\left(m^{2}+\varrho R\right)\phi^{2}\right] \; .
\label{309}
\eeq
Here $R$ is the Ricci scalar corresponding to AdS (and it is a 
negative constant, $R=-d(d+1)$) and $\varrho$ is an arbitrary coupling 
coefficient. The 
equation of motion generalizes to $\left[\nabla^{2} - (m^{2}+\varrho 
R)\right]\phi =0$. In 
this situation, Eqs.(\ref{9006''}, \ref{9006'''}) should 
include a dependence on $\varrho$, and thus be replaced by

\beq
\Delta_{\pm}(\varrho)=\frac{d}{2}\;\pm\;\nu(\varrho)\; ,
\label{7000}
\eeq
\beq
\nu(\varrho)=\sqrt{\frac{d^2}{4}\;+\;m^{2}+\varrho R}\; .
\label{7001}
\eeq
As before, irregular modes propagate for

\beq
\nu(\varrho)< 1\; .
\label{7002}
\eeq
Note that the Breitenlohner-Freedman bound Eq.(\ref{9034}) now reads

\beq
m^{2}+\varrho R\geq -\frac{d^{2}}{4}\; .
\label{7003}
\eeq
When performing the quantization of the non-minimally coupled scalar field 
in AdS, the energy of such a theory was defined in 
\cite{freedman}\cite{freedman8} as the conserved charge arising from the 
current which is obtained by contracting the stress-energy tensor with 
the Killing vector corresponding to time translations. With this 
definition, the energy is conserved, 
positive and finite for irregular modes propagating in the bulk when the 
following constraint is satisfied \cite{freedman}\cite{freedman8}

\beq
\varrho = \frac{1}{2}\;\frac{\Delta_{-}(\varrho)}
{1+2\Delta_{-}(\varrho)}\; .
\label{9041}
\eeq
Now, the Legendre transformation in \cite{witten2} can be performed for 
any values of $\varrho$, so that the 
constraint Eq.(\ref{9041}) should be imposed by hand. An additional 
difficulty was that such a prescription, where the leading non-local term 
of the action is selected {\it before} performing the Legendre 
transformation, does not work for $\nu =0$, due to the
presence of a logarithmic term in the generating functional \cite{our2}.

In order to solve these problems, the proposal in \cite{our3} was to 
introduce a modified formulation both in the bulk and in the the AdS/CFT 
correspondence. From the bulk point of view, it was performed a modified 
quantization where the `canonical' energy, which is constructed out of the 
Noether current corresponding to time displacements, is employed instead 
of the usual `metrical' one which is defined through the stress-energy 
tensor, as in \cite{freedman}\cite{freedman8}. The reason for this is 
that, as explained in \cite{our3}, the canonical energy is sensitive 
to the addition of boundary terms to the action, a property inherited 
from the Noether current, whereas the usual metrical one is 
not.\footnote{Recently, another
definition of the energy of the scalar field theory in AdS, which is
also sensitive to the boundary conditions, was
introduced in \cite{wald}.} This 
modified quantization gives rise to 
new constraints for the propagation of irregular modes (which make the 
canonical energy to be conserved, positive and finite) that come to 
replace Eq.(\ref{9041}). Such constraints depend on the particular 
boundary term which is added to the action, and many different examples 
were considered in \cite{our3}. On the other hand, the constraint 
Eq.(\ref{7002}), together with the Breitenlohner-Freedman bound (see 
Eqs.(\ref{9034}, \ref{7003})), remain unchanged, as expected.

From the AdS/CFT point of view, the proposal in \cite{our3} was to 
consider a modified prescription where the source $\phi_{0}$ in 
Eq.(\ref{9003}) is replaced by a more general one, which depends on 
the 
boundary conditions, or equivalently, on the boundary term which is added 
to the action, and could be a combination 
of both the field and its normal derivative. In addition, it was 
considered a generalized Legendre
transform prescription in which the Legendre transformation is performed 
on
the whole on-shell action, containing all local and non-local 
terms, instead of only on the leading non-local term,
as in the usual prescription. The generalized Legendre transformation 
contains all 
the information about the constraints arising from the quantization in the 
bulk, and it was shown in \cite{our3} that it solves all problems 
mentioned above regarding the usual formulation. In particular, the main 
goal in \cite{our3} was to show, for many different boundary terms added 
to the action, that the constraints for which the 
irregular modes propagate in the bulk when the canonical energy is 
considered instead of the metrical one, are the same for which the
divergent local terms of the on-shell action cancel out, and the 
generalized 
Legendre transformation interpolates between different 
conformal dimensions, namely $\Delta_{+}$ and $\Delta_{-}$. 

Motivated by the above detailed results in \cite{troost}, regarding the 
relation between 
unstable double-trace perturbations in the boundary, and bound states in 
the bulk, the purpose of this paper is to show that the action of the bulk 
theory governs, via the addition of boundary terms, the existence of a 
bound state in the bulk. We aim at computing, in both minimally and 
non-minimally coupled cases, which
boundary terms give rise to the existence of a bound state in the bulk,
and which ones do not. We also would like to incorporate relevant 
double-trace perturbations, and the description of bound states in the 
bulk and unstable theories at the boundary, into an extended 
formalism which is able to describe phenomena involving both the AdS/CFT 
duality and the quantization of the theory in the bulk. In this respect, 
the constraints computed in \cite{our3} for the propagation of irregular 
modes in the bulk will appear here again, this time in the role of the 
points where the relevant double-trace perturbations begin.

In Section 2, we introduce the 
background formalism regarding irregular modes, which will be the basis 
for the rest of this paper. With illustrative purposes, we will add to 
the action Eq.(\ref{301}) the simplest possible boundary term, which is 
quadratic in the field. We will consider the canonical energy of the 
theory for 
irregular modes propagating in the bulk, and, in the AdS/CFT context, we 
will analyze some subtleties regarding the Legendre transformation. Then, 
we will focus on a relevant double-trace perturbation at the border and, 
by also making use of the results in \cite{troost}, we will show that the 
boundary term in the action governs the existence of a bound 
state in the bulk, which will be related to an unstable perturbation at 
the boundary. A particular notorious example will be that of the usual 
action 
Eq.(\ref{301}), which contains no additional boundary term. We will 
show that it has an associated bound state. In general, in performing 
calculations we 
will pay a careful attention to the fact that relevant double-trace 
perturbations begin at the special points where irregular modes are 
allowed to propagate, as computed in \cite{our3}. 

In Section 3, we will consider the case of a 
non-minimally coupled scalar field, supplemented by a Gibbons-Hawking 
boundary term \cite{gibbons}, plus an additional, optional mass term at 
the 
boundary. We will show the existence of allowed
values of $\varrho$ (see Eq.(\ref{309})) for which there is no 
bound state in the bulk. In particular, we will show that a conformally 
coupled scalar field has no associated bound state.

In Section 4, we introduce, in both minimally and
non-minimally coupled cases, all remaining boundary
terms allowed by the variational principle, whose analysis allows to
perform additional consistency checks on
the formalism. In particular, we will reproduce, once again, the result
that
the usual action Eq.(\ref{301}), containing no additional boundary term,
has an associated bound state.

Finally, in Section 5, we revisit the Breitenlohner-Freedman bound, and
argue that the requirement for it to be satisfied does not suffice to
prevent tachyonic behavior
from existing in the bulk. It must be supplemented by additional    
conditions on the coefficients in the boundary terms of the action, i.e., 
the ones computed in Sections 2, 3 and 4. The reason for this is 
that the Breitenlohner-Freedman bound misses the part of the information 
in the action which is contained in the boundary terms. We also formulate 
our concluding remarks.

\section{Irregular Modes and Bound States}

In this paper, we will consider double-trace perturbations 
by a relevant operator ${\cal O}^{2}_{\alpha}\;$, where 
${\cal O}_{\alpha}$ has dimension $\Delta_{-}$ (see the Introduction for 
notation and details). But such a 
conformal operator is associated to irregular modes propagating in the 
bulk, and this suggests that, in order to get a complete understanding 
of such relevant double-trace perturbations, we should first carefully 
analyze the phenomenon of 
the propagation of irregular modes. In order to do this, we first 
introduce some background results which will be
extensively used throughout this paper. We will closely follow 
\cite{our3}, and also write 
parts of the formalism developed in such a reference in a more refined 
manner, which will be useful to our present purposes. In particular, we 
will focus on some subtleties regarding the Legendre transformation, which 
were 
not considered in \cite{our3}. Then, we will introduce a relevant 
double-trace perturbation at the boundary, and show that boundary terms in 
the action govern the existence of a bound state in the bulk. The 
non-minimally coupled case will be analyzed in the following section.

We begin by considering global coordinates in a $(d+1)$-dimensional AdS 
space. After 
setting the radius of $AdS_{d+1}$ equal to one, the metric 
reads

\beq
ds^{2} =
\frac{1}{cos^{2}\rho}\;(-d\tau^{2}+d\rho^{2}+sin^{2}\rho\;
d\Omega_{d}^{2})\quad (d\geq 2)\; ,
\label{39}
\eeq
where $d\Omega_{d}^{2}$ is the angular element, and $\rho$ and $\tau$ are 
the radial and time coordinates respectively. They satisfy

\ba 
&& \quad 0\leq\rho<\frac{\pi}{2}\qquad (d\geq 2)\; ,
\label{36}
\\
&& -\infty<\tau<\infty\; .
\label{35}
\ea
The above equation (which replaces $-\pi\leq\tau<\pi$) indicates that we 
are considering the universal 
covering space CAdS. This is done in order to avoid closed timelike 
curves (see for instance \cite{avis}). 

We consider the AdS space as foliated 
by $d$-dimensional surfaces $\partial{\cal 
M}_{\rho}$ of fixed radial coordinate $\rho$. Such surfaces 
are homeomorphic to
the boundary $\partial{\cal M}$ at $\rho\rightarrow\frac{\pi}{2}\;$. 
We refer
to $\partial{\cal M}_{\rho}$ as the boundary to the interior region
${\cal M}_{\rho}\;$. The limit $\rho\rightarrow\frac{\pi}{2}$ is to be 
taken 
only at the end of calculations. The surface forming
an outer normal vector to $\partial{\cal M}_{\rho}$ is given 
by

\beq
n_{\mu}\; =\;\frac{1}{cos\rho}\;\delta^{(\rho)}_{\mu}\; .
\label{40'}
\eeq

We first focus on the case of a minimally coupled scalar field. The 
non-minimally coupled case will be analyzed in the following section. For 
reasons to be clarified later, it will be relevant to our work to 
generalize the usual action Eq.(\ref{301}) by adding a surface term 
to it. There are different possible choices for such a surface term (see 
Section 4). The 
simplest one is as follows

\beq
I_{1}=I_{0} - \lambda_{1}\int_{\partial{\cal M}_{\rho}}
d^{d}y\;\sqrt{h}\;\phi^{2}\; ,
\label{9000}
\eeq
where $\lambda_{1}$ is a coefficient, 
and $h_{\mu\nu}$ is the induced metric.
The above action was not considered in \cite{our3}, but the calculations 
are analogous to the ones involving other surface terms. Some results we 
will find were already considered in \cite{our4}, but here we will 
give a more detailed account of the calculations, as this will be useful 
to our present purposes.

Following a procedure analogous to that in \cite{our3}, we compute the 
Noether current corresponding to time displacements (which are isometries 
of the background metric), and then, making use of the equation of 
motion, we find the following expression for the canonical energy

\beq
E_{1} = -\int d^{d}y\;\sqrt{g}\;\left[
\Theta^{\tau}_{\;\; \tau} -\;\lambda_{1}\;\nabla_{\mu}
\left(n^{\mu}\phi^{2}\right)\right]\; , 
\label{41}
\eeq
where the global minus sign is due to the `mostly plus' signature of 
the metric, the integration is carried out over the spatial coordinates, 
and

\beq
\Theta_{\mu\nu} = \partial_{\mu}\phi\;\partial_{\nu}\phi\;-\;\frac{1}{2}
\; g_{\mu\nu}\left(
g^{\alpha\beta}\partial_{\alpha}\phi\;\partial_{\beta}\phi\;
+m^{2}\phi^{2}\right)\; .
\label{739}
\eeq
It can be 
shown that $E_{1}$ is conserved, positive and finite for irregular 
modes 
propagating in the bulk only when Eq.(\ref{12000}), together with the 
following constraint \cite{our4}

\beq
\lambda_{1}=\frac{\Delta_{-}}{2}\; ,
\label{9001}
\eeq
are satisfied. This allows to perform a quantization of the scalar field 
in AdS, in a 
manner described in \cite{our3}, and which is analogous to the one 
considered in \cite{freedman}\cite{freedman8}. However, the details of 
such calculations are not relevant to our present purposes. The constraint 
Eq.(\ref{9001}), which is intimately related to the propagation of 
irregular modes in the bulk, will be of a fundamental importance in what 
follows. The reason for this is that we will consider 
perturbations at the border by relevant conformal operators, which 
correspond to irregular modes in the bulk (as $2\Delta_{-}\leq d$, see 
Eq.(\ref{9006''})).

Once we have shown how Eq.(\ref{9001}) arises when working in 
global coordinates and requiring the canonical 
energy to be conserved, positive and finite for irregular modes, let us 
show how the same constraint arises from AdS/CFT calculations. This takes 
us to consider the Euclidean 
representation of the $AdS _{d+1}$ (with radius equal to one) in 
Poincar\'e coordinates described by the
half space $x _{0}>0$, $x _{i} \in {\bf R}$ with metric

\beq
ds^{2}=\frac{1}{x _{0}^{2}} \sum_{\mu=0}^{d} dx^{\mu}dx^{\mu}.
\label{3000}
\eeq
The space will be considered as foliated by a
family of surfaces $x_{0}=\epsilon$ where we will formulate a 
boundary-value problem for the scalar field. As pointed out in 
\cite{freedman3}, the limit $\epsilon\rightarrow 0$ is to be taken 
only at the end of calculations. Note that, in this coordinates, the 
outward pointing unit 
normal vector is given by\footnote{Note that $n^{\mu}$ is not a
true vector, as its variation under an infinitesimal diffeomorphism 
includes an extra deviation term. See for instance the Appendix of 
\cite{pons}.}

\beq
n_{\mu} = \left(-\epsilon^{-1},{\bf 0}\right)\; .
\label{3005}
\eeq
In the Euclidean coordinates Eq.(\ref{3000}), the action Eq.(\ref{9000}) 
reads

\beq
I_{1}=I_{0} + \lambda_{1}\int
d^{d}x\;\sqrt{h}\;\phi^{2}_{\epsilon}\; ,
\label{3001}
\eeq
where $\phi_{\epsilon}$ is the value of the field at $x_{0}=\epsilon$, and

\beq
I_{0}=\frac{1}{2}\;\int d^{d+1} x \;\sqrt{g}\;
\left(  
g^{\mu\nu}\partial_{\mu}\phi\;\partial_{\nu}\phi\; +
\;m^{2}\phi^{2}\right) \; .
\label{3002}
\eeq
Here a Wick rotation has been performed. 

We wish to consider a boundary-value problem on the scalar field. Note 
that under an infinitesimal variation

\beq
\phi\rightarrow\phi +\delta\phi\; ,
\label{716}
\eeq
the action Eq.(\ref{3001}) transforms as follows

\beq
\delta_{\phi} I_{1}=\int d^{d}x\;\sqrt{h}\;\psi^{(1)}_{\epsilon}\;\delta
\phi_{\epsilon}\; ,
\label{717}
\eeq
where the absence of a bulk contribution is due to the equation of 
motion. Here $\psi^{(1)}$ is defined 
through

\beq
\psi^{(1)}=\partial_{n}\phi + 2\lambda_{1}\phi\; ,
\label{9003''}
\eeq
where $\partial_{n}\phi$ is the normal derivative of $\phi$
   
\beq
\partial_{n}\phi = n^{\mu}\partial_{\mu}\phi\; .
\label{718}
\eeq
From Eq.(\ref{717}), we conclude that the action is stationary under a 
Dirichlet boundary condition at $x_{0}=\epsilon$

\beq 
\delta\phi_{\epsilon}=0\; .
\label{3015}
\eeq
Integrating by parts, and making use of the equation of motion, $I_{1}$
can be written as the following pure-surface term

\beq
I_{1} = \frac{1}{2}\int
d^{d}x\;\sqrt{h}\;\phi_{\epsilon}\;\psi^{(1)}_{\epsilon}\; .
\label{3040}
\eeq
We will make use of the
solution to
the equation of motion which is regular at $x_{0}\rightarrow\infty$. It 
reads \cite{freedman3}

\beq
\phi(x) = \int\frac{d^{d}k}{\left(2\pi\right)^{d}}\;
e^{-i\vec{k}\cdot\vec{x}}\;
x_{0}^{\frac{d}{2}}\;a(\vec{k})\;K_{\nu}(kx_{0}),
\label{3043}
\eeq
where $\vec{x}=(x^{1},...,x^{d})$, $k = \mid\vec{k}\mid$,
$K_{\nu}$ is the modified Bessel function, and $\nu$ is given by
Eq.(\ref{9006'''}). 

We have just seen that, in the particular case of the action $I_{1}$, we 
are in 
the presence of a Dirichlet boundary-value problem which fixes 
$\phi_{\epsilon}$ at the boundary (see Eq.(\ref{3015})). This means that 
we should write $\psi^{(1)}_{\epsilon}$ in terms of 
the boundary data $\phi_{\epsilon}$. We find

\beq
\psi^{(1)}_{\epsilon}(\vec{k})=-\left[\frac{d}{2}\;+\nu-2\lambda_{1}-k
\epsilon\;\frac{K_{\nu+1}(k\epsilon)}{K_{\nu}(k\epsilon)}\right]
\phi_{\epsilon}(\vec{k})\; ,
\label{3045'}
\eeq
where $\phi_{\epsilon}(\vec{k})$ is the Fourier transform of
$\phi_{\epsilon}(\vec{x})$. Inserting the above equation into 
Eq.(\ref{3040}), we arrive at

\beq
I_{1}\left[\phi_{\epsilon}\right] = -\frac{1}{2}\int d^{d}x
\; d^{d}y\;\sqrt{h}\;\phi_{\epsilon}(\vec{x})
\;\phi_{\epsilon}(\vec{y})
\int\frac{d^{d}k}{\left(
2\pi\right)^{d}}\;e^{-i\vec{k}\cdot\left(
\vec{x}-\vec{y}\right)}\;\left[\frac{d}{2} + \nu -2\lambda_{1}-
k\epsilon\;\frac{K_{\nu+1}(k\epsilon)}{K_{\nu}(k\epsilon)}\right].  
\label{3047}
\eeq

In order to get the full information about the boundary CFT, we still need 
to 
compute the Legendre transform of $I_{1}$. We should follow a procedure 
analogous to that in \cite{our3}, and 
perform the 
generalized Legendre transformation on the expression 
Eq.(\ref{3047}). Even when such a 
procedure is correct, a more illuminating point of view
will arise by performing the Legendre transformation {\it before} writing 
$\psi^{(1)}_{\epsilon}$ in terms of the boundary data $\phi_{\epsilon}$, 
i.e. on Eq.(\ref{3040}), instead of Eq.(\ref{3047}).

We begin by writing Eq.(\ref{3040}) as

\beq
I_{1}[\phi_{\epsilon}] = \frac{1}{2}\int
d^{d}x\;\sqrt{h}\;\phi_{\epsilon}\;
\left(\psi^{(1)}_{\epsilon}[\phi_{\epsilon}]\right)\; 
,
\label{3040'}
\eeq
where the notation $\psi^{(1)}_{\epsilon}[\phi_{\epsilon}]$ 
explicitly indicates that the boundary data is $\phi_{\epsilon}$, and 
that $\psi^{(1)}_{\epsilon}$ must be written in terms of it. Note from 
Eq.(\ref{3045'}) the identity

\beq
\frac{\delta\psi^{(1)}_{\epsilon}}{\delta\phi_{\epsilon}}=
\frac{\psi^{(1)}_{\epsilon}}{\phi_{\epsilon}}\; .
\label{3041}
\eeq
So, let
${\tilde\phi}_{\epsilon}$ stand for the Legendre conjugate of
$\phi_{\epsilon}\;$. The generalized Legendre transformation, including 
all local and non-local terms in the on-shell action, is performed in 
momentum space for convenience. It reads

\beq
{\cal J}_{1}[\phi_{\epsilon},{\tilde\phi}_{\epsilon}]=\frac{1}{2}\int 
\frac{d^{d}k}{(2\pi)^{d}}\;\phi_{\epsilon}({\vec k})\;
\psi^{(1)}_{\epsilon}(-{\vec k})-\int 
\frac{d^{d}k}{(2\pi)^{d}}\;\phi_{\epsilon}({\vec k})\;
{\tilde\phi}_{\epsilon}(-{\vec k})\; .
\label{5100}
\eeq
Note that the above generalized Legendre transformation contains all local 
and 
non-local terms of the action, via Eq.(\ref{3045'}). Now, setting 
$0=\frac{\partial {\cal J}_{1}}{\partial\phi_{\epsilon}}\;$ (for fixed 
${\tilde\phi}_{\epsilon}$), we find

\beq
0=\frac{1}{2}\;\psi^{(1)}_{\epsilon}+\frac{1}{2}\;\phi_{\epsilon}
\;\frac{\delta\psi^{(1)}_{\epsilon}}{\delta\phi_{\epsilon}}-
{\tilde\phi}_{\epsilon}\; ,
\label{5101}
\eeq
and using Eq.(\ref{3041}), we arrive at

\beq
{\tilde\phi}_{\epsilon}=\psi^{(1)}_{\epsilon}\; .
\label{5102}
\eeq
This is the demonstration that $\phi_{\epsilon}$ and 
$\psi^{(1)}_{\epsilon}$ are Legendre conjugates, a result to which we 
will come back later. Note that we could have arrived 
to Eq.(\ref{5102}) also by performing the Legendre transformation on 
Eq.(\ref{3047}) instead of Eq.(\ref{3040'}), i.e. by following a procedure 
analogous to the one considered in \cite{our3}. But in such case, we 
should follow an indirect path, by first computing the relation between 
${{\tilde\phi}_{\epsilon}}$ and $\phi_{\epsilon}$, and then verifying 
that it is identical to that between $\psi^{(1)}_{\epsilon}$ and 
$\phi_{\epsilon}$ (see Eq.(\ref{3045'})). Summarizing, both procedures 
contain 
exactly the same information, as expected, but the above considered one 
is more 
compact and illuminating, as it contains Eq.(\ref{5102}) as a necessary 
intermediate result. Due to the relevance of Eq.(\ref{5102}) to 
our present purposes, the above detailed procedure is the one that we will
employ in 
this paper, in the context of double-trace perturbations. 

Now, 
introducing Eq.(\ref{5102}) into Eq.(\ref{5100}), we find the 
Legendre transform of $I_{1}$ (see Eq.(\ref{3040'})), which reads

\beq
{\tilde I}_{1}[\psi^{(1)}_{\epsilon}] = -\frac{1}{2}\int
d^{d}x\;\sqrt{h}\;\left(\phi_{\epsilon}[\psi^{(1)}_{\epsilon}]\right)
\;\psi^{(1)}_{\epsilon}\; ,
\label{5103}
\eeq
where we explicitly indicate through the notation 
$\phi_{\epsilon}[\psi^{(1)}_{\epsilon}]$ that, unlike the original 
functional Eq.(\ref{3040'}), now it is $\phi_{\epsilon}$ which must be 
written in terms of $\psi^{(1)}_{\epsilon}$. Introducing Eq.(\ref{3045'}) 
into the above equation, we find

\beq
{\tilde I}_{1}\left[\psi^{(1)}_{\epsilon}\right]=\frac{1}{2}\int
d^{d}x
\; d^{d}y\;\sqrt{h}\;\psi^{(1)}_{\epsilon}(\vec{x})
\;\psi^{(1)}_{\epsilon}(\vec{y})\int\frac{d^{d}k}{\left(
2\pi\right)^{d}}\;e^{-i\vec{k}\cdot\left(
\vec{x}-\vec{y}\right)}\;\frac{1}{\frac{d}{2} + \nu -2\lambda_{1}-
k\epsilon\;\frac{K_{\nu+1}(k\epsilon)}{K_{\nu}(k\epsilon)}}\; ,
\label{3048}
\eeq
which, together with Eq.(\ref{3047}), contains the information about the 
boundary dual theory. 

We consider here the case of $\nu$ a not integer value satisfying 
Eq.(\ref{12000}), which is the relevant one to our present purposes, as we 
are 
interested in analyzing the situations when both regular and irregular 
modes are allowed to propagate in the bulk.\footnote{Other values of 
$\nu$, such as $\nu 
>1$, integer or not, and $\nu =0$, can be 
considered following procedures analogous to the ones in \cite{our3}, but 
we will not analyze them here.} Expanding Eqs.(\ref{3047}, \ref{3048}) in 
powers of $\epsilon$, we find

\ba
I_{1}[\phi_{\epsilon}] &=& -\frac{1}{2}\int d^{d}x
\; d^{d}y\;\phi_{\epsilon}(\vec{x})
\;\phi_{\epsilon}(\vec{y})\;\epsilon^{-d}\nonumber\\
&&\qquad\times\int\frac{d^{d}k}{\left(
2\pi\right)^{d}}\;e^{-i\vec{k}\cdot\left(
\vec{x}-\vec{y}\right)}\;{\bigg [}\;(\Delta_{-}-2\lambda_{1})
- 2^{1-2\nu}
\;\frac{\Gamma (1-\nu)}{\Gamma
(\nu)}\; (k\epsilon)^{2\nu}+\;\cdots\; {\bigg ]}\; ,\nonumber\\
\label{3080}
\ea

\ba
{\tilde I}_{1}[\psi^{(1)}_{\epsilon}] &=&\frac{1}{2}\int d^{d}x
\; d^{d}y\;\psi^{(1)}_{\epsilon}(\vec{x})
\;\psi^{(1)}_{\epsilon}(\vec{y})\;\epsilon^{-d}\nonumber\\
&&\qquad\times\int\frac{d^{d}k}{\left(
2\pi\right)^{d}}\;e^{-i\vec{k}\cdot\left(
\vec{x}-\vec{y}\right)}\;\frac{1}{(\Delta_{-}-\;2\lambda_{1})
- 2^{1-2\nu}
\;\frac{\Gamma (1-\nu)}{\Gamma
(\nu)}\; (k\epsilon)^{2\nu}+\;\cdots}\; ,\nonumber\\
\label{3081}
\ea
where the dots stand for higher orders. Note that here the 
constraint Eq.(\ref{9001}) arises again, this time in a different 
context, as this is precisely the 
situation for which the divergent local term in Eqs.(\ref{3080}, 
\ref{3081}) vanishes. Let us first consider the case when Eq.(\ref{9001}) 
is not satisfied, i.e. when only regular modes are allowed to propagate 
in the bulk. Then, integrating over ${\vec k}$ we get 

\ba
I_{1}[\phi_{\epsilon}] &=&
\mbox{divergent local terms}
\nonumber\\
 && - \frac{\nu}{\pi^{\frac{d}{2}}}\;
\frac{\Gamma(\Delta_{+})}{\Gamma(\nu)}
\; \int d^{d}x \;
d^{d}y\;\phi_{\epsilon}(\vec{x})
\;\phi_{\epsilon}(\vec{y})\;\frac{\epsilon^{-2\Delta_{-}}}{|\;
\vec{x}-\vec{y}\;|^{2\Delta_{+}}}\;+\;\cdots,
\label{3090} \\ \nonumber\\
{\tilde I}_{1}[\psi^{(1)}_{\epsilon}] 
&=& \mbox{divergent local terms}
\nonumber\\
 && - \frac{\nu}{\pi^{\frac{d}{2}}}\;
\frac{1}{\left(\Delta_{-}-2\lambda_{1}\right)^{2}}
\;\frac{\Gamma(\Delta_{+})}{\Gamma(\nu)}
\; \int d^{d}x \;
d^{d}y\;\psi^{(1)}_{\epsilon}(\vec{x})
\;\psi^{(1)}_{\epsilon}(\vec{y})
\;\frac{\epsilon^{-2\Delta_{-}}}{|\;
\vec{x}-\vec{y}\;|^{2\Delta_{+}}}
\nonumber\\ && +\cdots,
\label{3091}
\ea
where the dots stand for higher orders. The non-local term in 
Eq.(\ref{3090}) was analyzed in \cite{freedman3}\cite{viswa1}, and the one 
in Eq.(\ref{3091}) differs from it only by a normalization coefficient. 
The limit $\epsilon\rightarrow 0$ is taken through

\beq
\lim _{\epsilon\rightarrow
0}\epsilon^{-\Delta_{-}}\;\phi_{\epsilon}(\vec{x})
= \phi_{0}(\vec{x}),
\label{3092}   
\eeq
which is the usual limit, and

\beq
\lim _{\epsilon\rightarrow
0}\epsilon^{-\Delta_{-}}\;\psi^{(1)}_{\epsilon}(\vec{x})
= \psi^{(1)}_{0}(\vec{x}),
\label{3093}
\eeq
which has the same form as Eq.(\ref{3092}). Both fields $\phi_{\epsilon}$ 
and $\psi^{(1)}_{\epsilon}$ exhibit the same behavior, due to the 
fact that only regular modes propagate, as Eq.(\ref{9001})
is not satisfied. Note that both functionals 
Eqs.(\ref{3090}, 
\ref{3091}) give rise to a boundary conformal operator with dimension 
$\Delta_{+}$, as expected.

A different picture emerges when Eq.(\ref{9001}) is satisfied, and 
irregular modes are allowed to propagate as well (we emphasize that, in 
this analysis, we 
are considering the case when Eq.(\ref{12000}) is also satisfied, which is 
the relevant one to our present purposes). In such a situation, the 
local divergent terms in Eqs.(\ref{3080}, \ref{3081}) vanish.
Note that $I_{1}$ still gives rise to the conformal dimension 
$\Delta_{+}$, as it reads

\beq
I_{1}[\phi_{\epsilon}] =
- \frac{\nu}{\pi^{\frac{d}{2}}}\;
\frac{\Gamma(\Delta_{+})}{\Gamma(\nu)}
\; \int d^{d}x \;
d^{d}y\;\phi_{\epsilon}(\vec{x})
\;\phi_{\epsilon}(\vec{y})\;\frac{\epsilon^{-2\Delta_{-}}}{|\;
\vec{x}-\vec{y}\;|^{2\Delta_{+}}}\;+\;\cdots.
\label{3090'}
\eeq
On the other hand, Eq.(\ref{3081}) is written

\beq
{\tilde I}_{1}[\psi^{(1)}_{\epsilon}]=
-\frac{1}{4\pi^{\frac{d}{2}}}
\;\frac{\Gamma(\Delta_{-})}{\Gamma(1-\nu)}\int d^{d}x \;
d^{d}y\;\psi^{(1)}_{\epsilon}(\vec{x}) 
\;\psi^{(1)}_{\epsilon}(\vec{y})
\;\frac{\epsilon^{-2\Delta_{+}}}{|\; 
\vec{x}-\vec{y}\;|^{2\Delta_{-}}}+\;\cdots,
\label{3101}
\eeq
and, instead of Eq.(\ref{3093}), we find the behavior

\beq
\lim _{\epsilon\rightarrow
0}\epsilon^{-\Delta_{+}}\;\psi^{(1)}_{\epsilon}(\vec{x})
= \psi^{(1)}_{0}(\vec{x}).
\label{5239}
\eeq
Note that, as expected, Eq.(\ref{3101}) gives rise to a conformal 
operator ${\tilde{\cal O}}$ with dimension $\Delta_{-}$, corresponding to 
irregular modes 
propagating in the bulk. The regular modes are accounted for by 
$I_{1}[\phi_{\epsilon}]$.

It will be useful to our present purposes to further elaborate on 
Eq.(\ref{3045'}). Note that, expanding in powers of $\epsilon$, we have

\beq
\psi^{(1)}_{\epsilon}=-\left(\Delta_{-}-2\lambda_{1} 
+\cdots\right)\phi_{\epsilon}\; ,
\label{5240}
\eeq
where the dots stand for higher orders. Using 
Eqs.(\ref{3092}, \ref{3093}), and taking the limit $\epsilon\rightarrow 
0$, we get

\beq
\psi^{(1)}_{0}=-\left(\Delta_{-}-2\lambda_{1}\right)\phi_{0}\; .
\label{5241}
\eeq
We will come back to the above equation later in this section.

At this point, it is interesting to note the analogy between 
Eqs.(\ref{3092}, \ref{5239}) 
and Eq.(\ref{8000}). It suggests that, in this formulation, $\phi_{0}$ and 
$\psi_{0}^{(1)}$ encode the information on $\beta$ and $\alpha$, 
respectively. We have just shown that 
$\phi_{\epsilon}$ and $\psi_{\epsilon}^{(1)}$ are Legendre conjugates (see 
Eq.(\ref{5102})), just as it happens to $\beta$ and $\alpha$. Note that, 
for regular 
modes, $\phi_{0}$ acts as the source (see Eq.(\ref{3090'})), as it happens 
to 
$\beta$ in Eq.(\ref{8000}). On the other hand, for irregular modes, 
it is $\psi_{0}^{(1)}$ that acts as the source (see Eq.(\ref{3101})), a 
role played by $\alpha$ in Eq.(\ref{8000}). A precise description of some 
aspects of 
Eq.(\ref{8000}) is perhaps more clearly seen in global 
coordinates Eq.(\ref{39}), when the quantization is performed (see, for 
instance, the discussion on 
regular and irregular modes in Section 3 of Ref.\cite{our3}, and 
references therein). But we have just shown that, in the present 
formulation, the information on the propagation of regular and irregular 
modes in the bulk is encoded in the Legendre conjugates $\phi_{\epsilon}$ 
and $\psi^{(1)}_{\epsilon}$, and we will make 
strong use of such result in what follows.

Now, we are led to analyze how to describe the 
perturbation at the boundary CFT by a relevant double-trace perturbation. 
The first thing to notice is that it should involve a conformal operator 
of dimension $\Delta_{-}$, as $2\Delta_{-}<d$. But we note from 
Eq.(\ref{3101}) that it corresponds to the conformal operator 
${\tilde{\cal O}}$, having $\psi_{0}^{(1)}$ as its source. The 
double-trace perturbation reads

\beq
W[{\tilde {\cal O}}]=\frac{f}{2}\;{\tilde{\cal O}}^{2}\; ,
\label{5243}
\eeq
which is the analogous to Eq.(\ref{8001}). Here $f$ is a 
coupling coefficient. But, as we have just pointed out, the source to 
${\tilde{\cal O}}$ is $\psi^{(1)}_{0}$, and this means that $\phi_{0}$ 
should be understood as its expectation value. Namely

\beq
\phi_{0}\equiv\left<{\tilde{\cal O}}\right>\; ,
\label{5244}
\eeq
which is the analogous to Eq.(\ref{8007}). From Eqs.(\ref{5243}, 
\ref{5244}), we can write

\beq
W[\phi_{0}]\equiv\frac{f}{2}\;\phi_{0}^{2}\; .
\label{5245}
\eeq

In this process, we have carefully identified the correct source 
for the conformal operator of dimension $\Delta_{-}$. But there is still 
another crucial observation to be made, which is that we have to consider 
the case when Eq.(\ref{9001}) is satisfied, and irregular modes, which 
correspond to the conformal dimension $\Delta_{-}$, are allowed to 
propagate. Note, also, that we should focus on the functional ${\tilde 
I}_{1}$, which is the one which gives rise to the conformal operator 
${\tilde{\cal O}}$ of dimension $\Delta_{-}$ (see Eq.(\ref{3101})). This 
means that the starting point is (see Eq.(\ref{5103}))

\beq
{\tilde I}_{1}
= -\frac{1}{2}\int
d^{d}x\;\sqrt{h}\;\phi_{\epsilon}
\;\psi^{(1)}_{\epsilon}\mid_{\lambda_{1} =\frac{\Delta_{-}}{2}}\; ,
\label{5290}
\eeq
where we have indicated that we are evaluating at the critical point 
Eq.(\ref{9001}) where irregular modes propagate, and from 
Eq.(\ref{9003''}) 
we have

\beq
\psi^{(1)}\mid_{\lambda_{1}
=\frac{\Delta_{-}}{2}}=\partial_{n}\phi + \Delta_{-}\phi\; .
\label{5248}
\eeq
We 
come back to the Legendre transformation Eq.(\ref{5100}), which is 
schematically written as

\beq
{\cal J}_{1}=\int
\frac{d^{d}k}{(2\pi)^{d}}\;\left(\frac{1}{2}\;\phi_{\epsilon}\;
\psi^{(1)}_{\epsilon}\mid_{\lambda_{1} =\frac{\Delta_{-}}{2}}-
\;\phi_{\epsilon}\;
{\tilde\phi}_{\epsilon}\right)\; .
\label{5270}
\eeq
From Eq.(\ref{5102}), ${\phi_{\epsilon}}$ and 
$\psi^{(1)}_{\epsilon}\mid_{\lambda_{1} =\frac{\Delta_{-}}{2}}$ are 
Legendre 
conjugates

\beq
{\tilde\phi}_{\epsilon}=\psi^{(1)}_{\epsilon}\mid_{\lambda_{1} 
=\frac{\Delta_{-}}{2}}\; .
\label{5271'}
\eeq

Now we perturb the boundary CFT by the relevant double-trace perturbation
Eq.(\ref{5243}). From Eqs.(\ref{5245}, \ref{5270}), this takes
${\cal J}_{1}$ to

\beq
{\cal J}_{1}\longrightarrow {\cal J}^{(f)}_{1}=\int
\frac{d^{d}k}{(2\pi)^{d}}\;\phi_{\epsilon}
\left[\frac{1}{2}\left(
\psi^{(1)}_{\epsilon}\mid_{\lambda_{1} 
=\frac{\Delta_{-}}{2}}+\; f\phi_{\epsilon}\right)-
{\tilde\phi}_{\epsilon}\right]\; ,
\label{5271}
\eeq
which can be written (see Eq.(\ref{9003''}))

\beq
{\cal J}^{(f)}_{1}=\int
\frac{d^{d}k}{(2\pi)^{d}}\;\left(\frac{1}{2}\;\phi_{\epsilon}\;
\psi^{(1)}_{\epsilon}\mid_{\lambda_{1} =\frac{\Delta_{-}}{2}+\frac{f}{2}}-
\;\phi_{\epsilon}\;
{\tilde\phi}_{\epsilon}\right)\; ,
\label{5270'}
\eeq
where

\beq
\psi^{(1)}\mid_{\lambda_{1}
=\frac{\Delta_{-}}{2}+\frac{f}{2}}=\partial_{n}\phi +
(\Delta_{-}+f)\phi\; .
\label{5274}
\eeq
Setting $0=\frac{\partial {\cal J}^{(f)}_{1}}{\partial\phi_{\epsilon}}\;$ 
(for fixed ${\tilde\phi}_{\epsilon}$), 
and using Eq.(\ref{3041}), we get

\beq
{\tilde\phi}_{\epsilon}=\psi^{(1)}_{\epsilon}\mid_{\lambda_{1}
=\frac{\Delta_{-}}{2}+\frac{f}{2}}\; ,
\label{5273}
\eeq
and inserting the above equation into Eq.(\ref{5270'}), we find

\beq
{\tilde I}^{(f)}_{1}
= -\frac{1}{2}\int
d^{d}x\;\sqrt{h}\;\phi_{\epsilon}
\;\psi^{(1)}_{\epsilon}\mid_{\lambda_{1} 
=\frac{\Delta_{-}}{2}+\frac{f}{2}}\; 
.
\label{5276}
\eeq

From the comparison between Eq.(\ref{5271'}) and Eq.(\ref{5273}), or 
between 
Eq.(\ref{5290}) and Eq.(\ref{5276}), we note that the effect of the 
relevant double-trace
perturbation Eq.(\ref{5243}) has been to introduce the replacement

\beq
\lambda_{1} =\frac{\Delta_{-}}{2}\longrightarrow \lambda_{1}
=\frac{\Delta_{-}}{2}+\frac{f}{2}\; .
\label{5252}
\eeq

Here is where we should include the sign of the coefficient $f$ in our
analysis. We know that positive $f$ corresponds to stable perturbations,
whereas negative $f$ corresponds to unstable ones \cite{witten7}. On the
other hand, the results in \cite{troost} indicate that the bulk theory
detects an unstable theory in the boundary through the existence of a
bound state. As pointed out in the Introduction, the results in 
\cite{troost} are based on a careful analysis of the spectrum of the 
radial wave-equation in Lorentzian Poincar\'e coordinates, and show 
the 
existence of a bound state with tachyonic behavior in a Minkowski slice 
when Eq.(\ref{8009}) is satisfied. Note that this relates the existence of 
a bound 
state to negative $f$ via Eq.(\ref{8002}). Thus, the author 
of \cite{troost} concludes that negative $f$ indicates the presence 
of a bound state in the bulk. But we note in Eq.(\ref{5252}) that negative 
$f$ is equivalent to the condition

\beq
\lambda_{1}<\frac{\Delta_{-}}{2}\; .
\label{5253}
\eeq 
In other words, given an action such as $I_{1}$ in Eq.(\ref{9000}), we
conclude that it will be associated
to a bound state, which is detected in a Minkowski slice in 
Lorentzian Poincar\'e 
coordinates, provided that Eq.(\ref{5253}) is satisfied. A
notorious particular case is that of $\lambda_{1} =0$, which satisfies
the condition Eq.(\ref{5253}), and thus has an associated
bound state. It corresponds to the usual action $I_{0}$ in
Eq.(\ref{301}), containing no additional
surface term. Allowed values of $\lambda_{1}$, for which there is no
associated bound state in the bulk, are the ones which do not satisfy
Eq.(\ref{5253}), i.e. 

\beq
\lambda_{1}\geq\frac{\Delta_{-}}{2}\;.
\label{52531}
\eeq

In this way, we have just demonstrated that the boundary terms in the
action govern the existence of a bound state in the bulk, which is 
detected in a Minkowski slice in Lorentzian Poincar\'e coordinates, 
in a manner described in \cite{troost}. Such a bound state is present even 
when the Breitenlohner-Freedman bound   
Eq.(\ref{9034}) is satisfied, which suggests that the last must be 
supplemented by Eq.(\ref{52531}). We will come back to this topic in 
Section 5. Notice that, in cases where a bound state is present, the 
addition of a proper boundary term to the action, as above described, 
should be required. 

As a last observation to be made, we point out that, 
by replacing $\lambda_{1}$ in Eq.(\ref{5241}) by the r.h.s. of 
Eq.(\ref{5252}) (i.e. $\lambda_{1}=\frac{\Delta_{-}}{2}+\frac{f}{2}$), we 
find

\beq
\psi^{(1)}_{0}= \; f\;\phi_{0}\; ,
\label{5255}
\eeq
which is the analog to Eq.(\ref{8002}).

\section{The Non-Minimally Coupled Case}

In this section, we focus on the non-minimally coupled case, where
we will show the existence of allowed values of the coupling coefficient
$\varrho$ (see Eq.(\ref{309})) for which there is no bound state in the
bulk. Notice that, in this situation, we should employ Eq.(\ref{7000}) 
instead of Eq.(\ref{9006''}).

We begin by considering the following action in global coordinates   
Eq.(\ref{39})

\beq
{\cal I}_{1}={\cal I}_{0}+\varrho\int_{\partial{\cal M}_{\rho}}
d^{d}y\;\sqrt{h}\; K\;\phi^{2}-\sigma\int_{\partial{\cal M}_{\rho}}
d^{d}y\;\sqrt{h}\;\phi^{2}\; ,
\label{6000}
\eeq
where ${\cal I}_{0}$ is given by Eq.(\ref{309}), $K$ is the trace of 
the extrinsic curvature, and $\sigma$ is a coefficient. The first 
boundary term is
the natural extension of the Gibbons-Hawking term \cite{gibbons}, which is
needed in order to have a well-defined variational principle under
variations of the metric.\footnote{When performing a variation of the
metric, the action ${\cal I}_{0}$ in Eq.(\ref{309}) results to be
stationary only after
the metric and certain of its normal derivatives are fixed at the
boundary. It can be shown that the addition of the Gibbons-Hawking term
accounts for the terms of the variation containing derivatives of the
metric.} The last term in Eq.(\ref{6000}) has the form of a mass term 
which is added at the boundary. We include it for completeness, as it does 
not spoil the property of having a well-defined variational principle 
under variations of the metric. In the particular case $m=\sigma=0$, 
and when $\varrho$ satisfies

\beq
\varrho=\frac{d-1}{4d}\; ,
\label{6001}
\eeq
both the bulk and boundary terms in Eq.(\ref{6000}) are Weyl-invariant
(see for instance \cite{saharian} for a recent treatment). 

The canonical
energy of the theory Eq.(\ref{6000}) was computed in \cite{our3} in the 
particular case $\sigma=0$. Here we extend the results in \cite{our3} 
to arbitrary $\sigma$. It can be shown that the canonical energy is
conserved, positive and finite for irregular modes propagating in the bulk
only when the following constraint is satisfied

\beq
\varrho+\frac{\sigma}{d} =\frac{\Delta_{-}(\varrho)}{2d}\; ,
\label{6002}
\eeq
which comes to replace the usual constraint Eq.(\ref{9041}). It has
solutions

\beq
\varrho^{\pm}=\frac{d-1}{8d}\;
\left[1\; 
-\frac{8\sigma}{d-1}\;\pm\sqrt{1+
\left(\frac{4}{d-1}\right)^{2}\left[m^{2}+(d+1)\sigma
\right]}\right]\; , 
\label{151}
\eeq
which should be supplemented by the reality condition

\beq
m^{2}+(d+1)\sigma\geq-\left(\frac{d-1}{4}\right)^{2}\; .
\label{159}
\eeq
Note that, when $m=\sigma=0$, $\varrho^{-}$ vanishes, whereas 
$\varrho^{+}$ reduces to the conformal value Eq.(\ref{6001}). This could 
be considered as a check on the formalism.

In the case $\sigma=0$, it was shown in \cite{our3} that, as expected, 
the constraint
Eq.(\ref{6002}) arises again from AdS/CFT calculations, this time in the
role of the condition for
the divergent local terms in the on-shell action to vanish, and the
generalized Legendre transformation to interpolate between
different conformal dimensions $\Delta_{+}(\varrho)$ and
$\Delta_{-}(\varrho)$ (see Eq.(\ref{7000})). Such result is 
straightforwardly extended to arbitrary $\sigma$. Note that, in 
Euclidean Poincar\'e coordinates Eq.(\ref{3000}), ${\cal I}_{1}$ reads

\beq
{\cal I}_{1}={\cal I}_{0}-\varrho\int
d^{d}x\;\sqrt{h}\; K_{\epsilon}\;\phi^{2}_{\epsilon}+\sigma\int
d^{d}x\;\sqrt{h}\;\phi^{2}_{\epsilon}\; ,
\label{6000'}
\eeq
where

\beq
{\cal I}_{0}=\frac{1}{2}\;\int d^{d+1} x \;\sqrt{g}\;
\left[
g^{\mu\nu}\partial_{\mu}\phi\;\partial_{\nu}\phi\; +
\;\left(m^{2}+\varrho R\right)\phi^{2}\right] \; .
\label{309'}
\eeq
Since in Euclidean Poincar\'e
coordinates the trace of the extrinsic curvature satisfies
$K=-d$, we note that the AdS/CFT calculations for ${\cal I}_{1}$ are
analogous to the ones performed in the previous section in the minimally 
coupled case. All
we
need to do is to perform the replacements $\lambda_{1}\rightarrow\varrho
d +\sigma$,
$\nu\rightarrow\nu(\varrho)$ and
$\Delta_{\pm}\rightarrow\Delta_{\pm}(\varrho)$ (see Eqs.(\ref{7000},
\ref{7001})). We
find that the analogous to Eq.(\ref{5253}) reads

\beq
\varrho+\frac{\sigma}{d}<\frac{\Delta_{-}(\varrho)}{2d}\; ,
\label{6003}
\eeq
which is the condition for a bound state to exist in the bulk, and is
related to the presence of an unstable double-trace perturbation
at the boundary.

But the above condition, together with Eq.(\ref{7002}), still needs to be
solved for $\varrho$, a
further step which was not needed in the minimally coupled case. We first
note that the solution to Eq.(\ref{7002}) reads

\beq
\varrho>\frac{1}{4d(d+1)}\left[(d+2)(d-2)+4m^{2}\right]\; ,
\label{6004}
\eeq
which, together with the reality condition Eq.(\ref{159}), should always
be
required, as the case Eq.(\ref{7002}) is the
relevant one to our analysis. In fact, it will be useful to consider the
following condition
   
\beq
\varrho>\frac{1}{16d(d+1)}\left[4(d+2)(d-2)-(d-1)^{2}-16(d+1)
\sigma\right]\; ,
\label{9860}
\eeq
which is obtained from Eqs.(\ref{159}, \ref{6004}). Now, solving
Eq.(\ref{6003}) and using Eq.(\ref{9860}), we find that, for $d\geq 3$, a
bound state exists in the bulk only when the following conditions are
simultaneously satisfied

\beq
m^{2}+(d+1)\sigma <\frac{d}{4}\qquad\mbox{and}\qquad
\varrho^{+}<\varrho<\frac{1}{4}-\frac{\sigma}{d}\qquad\qquad\qquad 
(d\geq 3)\; .
\label{9861}
\eeq
Allowed values of $\varrho$, for which there is
no associated bound state, are the ones which do not satisfy the above   
conditions, i.e.

\ba 
m^{2}+(d+1)\sigma <\frac{d}{4}\; : &&
\varrho\leq\varrho^{+}\;\;\mbox{or}\;\;\varrho\geq\frac{1}{4}-
\frac{\sigma}{d}\;
,\nonumber\\
m^{2}+(d+1)\sigma\geq\frac{d}{4}\; :&& \mbox{any}
\;\varrho\quad\qquad\qquad\qquad\qquad (d\geq 3)\; .
\label{9862}
\ea
We emphasize that the above conditions should be required
simultaneously with Eq.(\ref{6004}), together with the reality
condition Eq.(\ref{159}), and the Breitenlohner-Freedman bound
Eq.(\ref{7003}). The remarkable result that Eqs.(\ref{9862}) supplement
the Breitenlohner-Freedman bound is far from trivial, and we
will come back to this topic in Section 5.

It is interesting to note that, in the notorious example of a
conformally coupled scalar field, supplemented by a Gibbons-Hawking  
term, there is no associated bound state, as
in such case $\varrho^{+}$ equals the conformal value Eq.(\ref{6001}), and
we have just seen that $\varrho^{+}$ is in the range of allowed values for
$\varrho$ (see Eqs.(\ref{9862})). In fact, as pointed out before, the case
$m=\sigma =0,\;\varrho=\varrho^{+}$ corresponds to an unperturbed
situation, where irregular modes are allowed to propagate.

To close this section, we extend the conditions Eqs.(\ref{9861}) for a
bound
state to exist in the bulk, which hold for $d\geq 3$, to the case $d=2$,
where a
bound state exists in the bulk only when the following conditions are
satisfied

\ba
m^{2}+3\sigma <\frac{1}{2}\; :&&\varrho<\varrho^{-}
\;\;\mbox{or}\;\;\varrho^{+}<\varrho<\frac{1}{4}-\frac{\sigma}{2}\; 
,\nonumber\\
m^{2}+3\sigma
\geq\frac{1}{2}\; :&&\varrho<\varrho^{-}\qquad\qquad
\qquad\qquad\qquad\quad (d=2)\; .
\label{9863}
\ea
Allowed values of $\varrho$, for which there is
no associated bound state, are the ones which do not satisfy
Eqs.(\ref{9863}), i.e.

\ba
m^{2}+3\sigma <\frac{1}{2}\; :&&\varrho^{-}
\leq\varrho\leq\varrho^{+}\;\;\mbox{or}
\;\;\varrho\geq\frac{1}{4}-\frac{\sigma}{2}\; ,\nonumber\\   
m^{2}+3\sigma\geq\frac{1}{2}\; :&&\varrho\geq\varrho^{-}\qquad\qquad   
\qquad\qquad\qquad\quad (d=2)\; .
\label{9864}
\ea
As in the case $d\geq 3$, the above conditions must be supplemented by
Eqs.(\ref{7003}, \ref{159}, \ref{6004}).

\section{Other Boundary Terms}

In the previous sections, we have illustrated our proposal that
boundary
terms in the action govern the existence of bound states in the
bulk, by considering the examples in Eqs.(\ref{3001}, \ref{6000'}) (in
Euclidean coordinates). Now, we
would like to analyze how the previous results are
modified when other boundary terms are considered. This will also allow
us to perform additional checks on the formalism.

We should
first notice that it is not possible to add any arbitrary boundary term to
the actions Eqs.(\ref{3002}, \ref{309'}). The reason for this is that we
should only consider situations where the variational principle is  
well-defined. In the minimally coupled case, it is possible to verify, by
direct inspection, that, apart from Eq.(\ref{3001}), we are left with only
two
additional possibilities, which in Euclidean coordinates Eq.(\ref{3000})
read \cite{our4}\cite{our3}

\beq
I_{2}=I_{0} + \lambda_{2}\int
d^{d}x\;\sqrt{h}\;\left(\partial_{n}\phi_{\epsilon}\right)^{2}\; ,
\label{4003}
\eeq
and

\beq
I_{3}=I_{0} + \lambda_{3}\int
d^{d}x\;\sqrt{h}\;\phi_{\epsilon}^{2}-\int
d^{d}x\;\sqrt{h}\;\phi_{\epsilon}\;\partial_{n}\phi_{\epsilon}\; ,
\label{4042}
\eeq
where $\lambda_{2}$ and $\lambda_{3}$ are arbitrary coefficients, and   
$I_{0}$ is given by Eq.(\ref{3002}).\footnote{There is a curious feature 
about Eqs.(\ref{4003},
\ref{4042}), shown
in \cite{our3}, which is the fact that, whereas the on-shell action
$I_{1}$ gives rise to the boundary CFT of conformal dimension
$\Delta_{+}$, and its Legendre transform ${\tilde I}_{1}$ corresponds to 
$\Delta_{-}$ (see Eqs.(\ref{3090'}, \ref{3101})), when considering $I_{2}$
and $I_{3}$ we find an `inverted' situation, where the original generating
functional corresponds to the conformal dimension $\Delta_{-}$, and the
Legendre transformed one is associated to $\Delta_{+}$ (this
`inversion' phenomenon is also found in the non-minimally coupled case
\cite{our3}). At this point, we do not know if such an `inversion'
phenomenon is, or not, associated to any property of the boundary
CFT.} Under
the variation Eq.(\ref{716}) we
have\footnote{With illustrative purposes, we give here an example of a
one-parameter family of boundary terms which is not allowed. Consider the
action $I'=I_{0}+\gamma\int
d^{d}x\;\sqrt{h}\;\phi_{\epsilon}\;\partial_{n}\phi_{\epsilon}$, where
$\gamma$ is an arbitrary coefficient, and $I_{0}$ is given by
Eq.(\ref{3002}). Then, under the variation Eq.(\ref{716}) we have
$\delta_{\phi} I'=\int
d^{d}x\;\sqrt{h}\left[(1+\gamma)\;\partial_{n}\phi_{\epsilon}
\;\delta\phi_{\epsilon}+\gamma\;\phi_{\epsilon}\;
\delta(\partial_{n}\phi_{\epsilon})\right]$. This result is to be
contrasted with Eqs.(\ref{717}, \ref{7700}). Notice that, in order for the
action to be stationary, we should fix both $\phi_{\epsilon}$ and
$\partial_{n}\phi_{\epsilon}$ at the border,
$\delta\phi_{\epsilon}=\delta(\partial_{n}\phi_{\epsilon})=0$. From the
AdS/CFT point
of view, the source for the boundary conformal operator is ill-defined,
unlike what happens to Eqs.(\ref{717}, \ref{7700}), where the sources are
$\phi_{\epsilon}$, $\psi^{(2)}_{\epsilon}$ and $\psi^{(3)}_{\epsilon}$
respectively
\cite{our4}\cite{our3}. This
situation is analogous
to the one found in the case of the Einstein-Hilbert action, whose
variation requires to fix at the border both the metric and its
derivatives. In such case, the requirement to have a well-defined
variational principle is satisfied through the addition of the
Gibbons-Hawking boundary term \cite{gibbons}, which accounts for the
derivatives of the
metric. Notice that, in the above case, the only possible choice for the
coefficient $\gamma$ is $\gamma =-1$, where the variation reduces to
$\delta_{\phi} I'=-\int
d^{d}x\;\sqrt{h}\;\phi_{\epsilon}\;
\delta(\partial_{n}\phi_{\epsilon})$. Here the action is stationary under
a Neumann boundary condition which fixes $\partial_{n}\phi_{\epsilon}$ at
the
boundary, $\delta(\partial_{n}\phi_{\epsilon})=0$. Note that the allowed
Neumann situation $\gamma=-1$ is a
particular case of Eq.(\ref{4042}) with $\lambda_{3}=0$.}

\beq
\delta_{\phi} I_{2}=\int
d^{d}x\;\sqrt{h}\;\partial_{n}\phi_{\epsilon}\;\delta
\psi^{(2)}_{\epsilon}\qquad\qquad\qquad
\delta_{\phi} I_{3}=-\int
d^{d}x\;\sqrt{h}\;\phi_{\epsilon}\;\delta
\psi^{(3)}_{\epsilon}\; ,
\label{7700}
\eeq
where

\beq
\psi^{(2)}=\phi + 2\lambda_{2}\;\partial_{n}\phi\qquad\qquad\qquad
\psi^{(3)}=\partial_{n}\phi-2\lambda_{3}\phi\; .
\label{7701}
\eeq
So, $I_{2}$ and $I_{3}$ are stationary under the mixed boundary conditions

\beq
\delta\psi^{(2)}_{\epsilon}=0\; ,
\label{7702}
\eeq
and

\beq
\delta\psi^{(3)}_{\epsilon}=0\; ,
\label{7703}
\eeq
respectively. In particular, this means that, after taking the limit
$\epsilon\rightarrow 0$ through a proper rescaling,
$\psi^{(2)}_{\epsilon}$ and $\psi^{(3)}_{\epsilon}$ become the
sources for the corresponding boundary conformal operators
\cite{our4}\cite{our3}.

In the case of $I_{2}$, it can be shown that the corresponding canonical 
energy is conserved, positive and finite for irregular
modes propagating in the bulk, or equivalently, that the divergent local
terms of the on-shell action cancel out and the Legendre transformation
interpolates between conformal dimensions $\Delta_{+}$ and $\Delta_{-}$,
only when Eq.(\ref{12000}), together with the constraint \cite{our3}

\beq
\lambda_{2}=\frac{1}{2\Delta_{-}}\; ,
\label{3401}
\eeq
are satisfied. Performing calculations analogous to those previously
detailed
in the case of $I_{1}$, we
find that $\psi^{(2)}_{\epsilon}$ and $\partial_{n}\phi_{\epsilon}$ are
Legendre conjugates, i.e.

\beq
{\tilde\psi}^{(2)}_{\epsilon}=\partial_{n}\phi_{\epsilon}\; .
\label{3400}
\eeq
Here $\psi^{(2)}_{\epsilon}$ and $\partial_{n}\phi_{\epsilon}$ are the
sources for the conformal operators of dimensions $\Delta_{-}$ and
$\Delta_{+}$ respectively, so that a relevant double-trace perturbation
at the boundary can be written

\beq
W[\partial_{n}\phi_{0}]
\equiv\frac{f}{2}\left(\partial_{n}\phi_{0}\right)^{2}\; ,
\label{3500}
\eeq
and introduces the replacement

\beq
\lambda_{2}=\frac{1}{2\Delta_{-}}\longrightarrow\lambda_{2}=
\frac{1}{2\Delta_{-}}+\frac{f}{2}\; .
\label{3501}
\eeq
Bound states exist in the bulk only when the following condition
is satisfied

\beq
\lambda_{2}<\frac{1}{2\Delta_{-}}\; .
\label{3402}
\eeq
Allowed values of $\lambda_{2}$, for which there is no bound state in the
bulk, are the ones in the range

\beq
\lambda_{2}\geq\frac{1}{2\Delta_{-}}\; .
\label{3403}
\eeq
A notorious example is that of the usual action $I_{0}$ in
Eq.(\ref{301}), containing no additional
surface term, which corresponds to the case $\lambda_{2}=0$ (see
Eq.(\ref{4003})), so that
it is
not in the range Eq.(\ref{3403}), and has an associated bound state. The
same
result has already been found when considering
$I_{1}$, and the fact that both analyses, involving $I_{1}$ or $I_{2}$, 
give rise to the same result, could be considered as a consistency  
check on the formalism.

Now, in the case of $I_{3}$, the constraint for which the canonical
energy is conserved, positive and finite for irregular modes propagating
in the bulk, and the divergent local terms in the on-shell action cancel
out, making the Legendre transformation to interpolate between conformal
dimensions $\Delta_{+}$ and $\Delta_{-}$, is given by \cite{our4}

\beq
\lambda_{3}=-\frac{\Delta_{-}}{2}\; .
\label{3404}
\eeq
It can be shown that $\psi^{(3)}_{\epsilon}$ and $-\phi_{\epsilon}$ are
Legendre conjugates, i.e.

\beq
{\tilde\psi}^{(3)}_{\epsilon}=-\phi_{\epsilon}\; .
\label{3405}
\eeq
Here $\psi^{(3)}_{\epsilon}$ and $-\phi_{\epsilon}$ are sources for the
conformal operators of dimensions $\Delta_{-}$ and $\Delta_{+}$
respectively, so that a
relevant double-trace perturbation is of the form

\beq
W[-\phi_{0}]
\equiv\frac{f}{2}\left(-\phi_{0}\right)^{2}\; ,
\label{3509}
\eeq
and performs the replacement

\beq
\lambda_{3}=-\frac{\Delta_{-}}{2}\longrightarrow\lambda_{3}=
-\frac{\Delta_{-}}{2}+\frac{f}{2}\; .
\label{3508}
\eeq
Bound states exist in the bulk only for $\lambda_{3}$ satisfying

\beq
\lambda_{3}<-\frac{\Delta_{-}}{2}\; .
\label{3406}
\eeq
Allowed values of $\lambda_{3}$, for which there is no bound state in the
bulk, are the ones in the range

\beq
\lambda_{3}\geq-\frac{\Delta_{-}}{2}\; .
\label{3407}
\eeq

Finally, in the non-minimally coupled case, we should take into account
that, in order to have a well-defined variational
principle under variations of the metric, any expression for the
action should contain a Gibbons-Hawking
term, as in Eq.(\ref{6000'}). It can be verified that, apart
from Eq.(\ref{6000'}), we are left with only one additional possibility,
namely

\beq
{\cal I}_{2}={\cal I}_{0} - \varrho\int
d^{d}x\;\sqrt{h}\;K_{\epsilon}\;\phi_{\epsilon}^{2}+
\sigma\int
d^{d}x\;\sqrt{h}\;\phi_{\epsilon}^{2}
-\int
d^{d}x\;\sqrt{h}\;\phi_{\epsilon}\;\partial_{n}\phi_{\epsilon}\; ,
\label{7704}
\eeq
where ${\cal I}_{0}$ is given by Eq.(\ref{309'}), and the first two
boundary terms are as in Eq.(\ref{6000'}). It can be shown that the last
surface term does not spoil the property of
having a
well-defined variational principle under variations of the metric.

The case $\sigma =0$ was
considered in \cite{our3}, but here we will extend such results to the
case of arbitrary $\sigma $. It can be shown that the canonical
energy is conserved, positive and finite for irregular modes propagating
in the bulk only when the following constraint is satisfied

\beq
\varrho+\frac{\sigma}{d} =-\frac{\Delta_{-}(\varrho)}{2d}\; ,
\label{7708}
\eeq
which is to be contrasted with Eq.(\ref{6002}). It has solutions

\beq
{\tilde\varrho}^{\pm}=\frac{3d+1}{8d}\;
\left[-1-\frac{8\sigma}{3d+1}
\pm\sqrt{1+\left(\frac{4}{3d+1}\right)^{2}\left[m^{2}+(d+1) 
\sigma\right]}\right]\; .
\label{7709}
\eeq
Notice the reality condition

\beq
m^{2}+(d+1)\sigma\geq-\left(\frac{3d+1}{4}\right)^{2}\; ,
\label{7710}
\eeq
which should always be required, together with Eq.(\ref{6004}). As
expected, the constraint Eq.(\ref{7708}) arises again from AdS/CFT
calculations, in the role of the condition for the divergent local terms
in the
on-shell action to vanish, and the Legendre transformation to interpolate
between different conformal dimensions $\Delta_{+}(\varrho)$ and
$\Delta_{-}(\varrho)$.

By performing calculations analogous to the ones in the previous cases,
we find that the condition

\beq
\varrho+\frac{\sigma}{d}<-\frac{\Delta_{-}(\varrho)}{2d}\; ,
\label{7711}
\eeq  
corresponds to the existence of a bound state in the bulk, in the sense
of \cite{troost}. Such condition is related to the presence of an unstable
double-trace perturbation at the boundary. Solving for $\varrho$, we find
the following solution

\ba
m^{2}+(d+1)\sigma<-\frac{1}{16}\left[(3d+1)^{2}-(d+5)^{2}\right]\; :&&
\varrho <{\tilde\varrho}^{+}\; ,\nonumber\\
m^{2}+(d+1)\sigma\geq
-\frac{1}{16}\left[(3d+1)^{2}-(d+5)^{2}\right]\; :&&\varrho
<-\frac{d-2}{4d}-\frac{\sigma}{d}\; .
\label{7712}
\ea
Allowed values of $\varrho$, for which there is
no associated bound state, are the ones which do not satisfy the above
conditions. This gives

\ba
m^{2}+(d+1)\sigma<-\frac{1}{16}\left[(3d+1)^{2}-(d+5)^{2}\right]
\; :&&
\varrho \geq{\tilde\varrho}^{+}\; ,\nonumber\\
m^{2}+(d+1)\sigma\geq
-\frac{1}{16}\left[(3d+1)^{2}-(d+5)^{2}\right]\; :&&
\varrho\geq-\frac{d-2}{4d}-\frac{\sigma}{d}\; .
\label{7713}
\ea 
We emphasize that the above conditions should be required simultaneously
with Eqs.(\ref{7003}, \ref{6004}, \ref{7710}).

Note that, in the notorious
particular case of a conformally coupled scalar field, where $m=\sigma=0$
and
$\varrho$ is given by Eq.(\ref{6001}), the conditions Eqs.(\ref{7713})
show that there is no bound state in the bulk, as it happened in the case
of ${\cal I}_{1}$ (see Eq.(\ref{6000'})). However, there is a fundamental
difference, because,
whereas for ${\cal I}_{1}$ the conformally coupled case corresponds to an
unperturbed situation where irregular modes are allowed to propagate in
the bulk, in the case of ${\cal I}_{2}$ it is associated to a (stable)
non-zero double-trace perturbation at the boundary. This result is
not surprising when we note that the last boundary term in Eq.(\ref{7704})
breaks the Weyl-invariance of Eq.(\ref{6000'}).

\section{Breitenlohner-Freedman Bound Revisited}

In this work, we have argued that 
coefficients in the boundary terms in the action are sensitive to 
the perturbation at the boundary CFT by a relevant double-trace 
operator (see for 
instance Eqs.(\ref{5252}, \ref{3501}, \ref{3508})), and govern the 
existence 
of a bound 
state 
in the bulk. The relation was made precise by using the proposal in 
\cite{troost} that 
unstable theories at the boundary are detected by the presence of such a 
bound state in the bulk. In all calculations, we have also made strong use 
of the 
formalism in \cite{witten7}\cite{our3}. In particular, we have paid a
careful attention to the fact that relevant double-trace perturbations are 
constructed out of a conformal operator of dimension $\Delta_{-}$. This 
means that we have to identify the correct source and generating 
functional for the conformal 
operator, and introduce the perturbations at the special points at which 
irregular modes are allowed to propagate (see Eqs.(\ref{9001}, \ref{6002}, 
\ref{3401}, \ref{3404}, \ref{7708})). From the bulk point of view, such 
special points 
arise from the requirement for the canonical energy to be conserved, 
positive and finite for irregular modes propagating in the bulk 
\cite{our3}. From AdS/CFT calculations, they play the role of the 
conditions for the divergent local terms in the on-shell action to vanish, 
and the generalized Legendre transform to interpolate 
between different conformal 
dimensions $\Delta_{+}$ and $\Delta_{-}$ \cite{our3}. 

Throughout this paper, we have 
considered many different allowed boundary terms in the 
action (see Eqs.(\ref{3001}, \ref{6000'}, \ref{4003}, \ref{4042}, 
\ref{7704})). By 
proposing such boundary terms to be the objects involved in the connection 
between unstable double-trace 
perturbations at the boundary and bound states in the bulk, we were able 
to compute explicit conditions on the coefficients of the boundary terms 
in the action for which we expect a bound state to exist in the bulk (see 
Eqs.(\ref{5253}, \ref{3402}, \ref{3406})). In the non-minimally coupled 
case, and when the action is supplemented by a Gibbons-Hawking term, this 
also gave rise to `forbidden' values of the coupling 
coefficient to the metric (see Eqs.(\ref{9861}, \ref{9863}, \ref{7712})). 

Notorious particular examples were also considered. For instance, we have 
shown that the usual action Eq.(\ref{301}), containing no 
additional boundary terms, is associated to the existence of a bound state 
in the bulk. Such result was found by considering independent analyses 
involving either actions $I_{1}$ or $I_{2}$ (see Eqs.(\ref{3001}, 
\ref{4003})). This could be considered as a consistency check on the 
formalism. Another notorious example has been that of a conformally 
coupled scalar field, supplemented by a Gibbons-Hawking term, to which we 
have shown that there is no associated bound state.

To close this paper, we come back to the result in 
\cite{troost} that tachyonic behavior in the bulk exists even when the 
Breitenlohner-Freedman bound (see Eqs.(\ref{9034}, \ref{7003})) is 
satisfied. We argue here that this happens so, because the 
Breitenlohner-Freedman bound should be supplemented by additional 
conditions involving the coefficients on the boundary terms in the action. 
Such conditions depend on the particular boundary term which is added to 
the action, and are given by Eqs.(\ref{52531}, \ref{3403}, \ref{3407}) 
(in the minimally coupled case), and Eqs.(\ref{9862}, \ref{9864}, 
\ref{7713}) (in 
the non-minimally coupled case, supplemented by a Gibbons-Hawking term, 
where we should simultaneously require the condition Eq.(\ref{6004}), 
together with the reality conditions Eq.(\ref{159}) for ${\cal I}_{1}$ or 
Eq.(\ref{7710}) for ${\cal I}_{2}$).
The reason why the 
requirement to satisfy the Breitenlohner-Freedman bound is not enough to 
prevent tachyonic behavior from existing is that such a bound misses a 
part of the information contained in the action, namely the one included 
in the boundary terms, and which is related to the boundary conditions on 
the field. 

The fact that some results are modified or generalized when the 
boundary terms in the action are taken into account is not surprising. 
See, for instance, the replacement of Eq.(\ref{9041}) by Eq.(\ref{6002}) 
or Eq.(\ref{7708}).

It would be interesting to investigate if any further information arising 
from the boundary terms in the action can be obtained by performing 
additional related calculations in Lorentzian Poincar\'e metric.

\section{Acknowledgments}

It is a pleasure to thank Prof. J. A. Helay\"el-Neto for interesting 
discussions and a critical reading of the manuscript. This paper was 
supported by CLAF/CNPq.

\end{document}